\documentclass[epj]{svjour}
\usepackage{graphics}
\usepackage{amsmath,amssymb, dsfont}
\usepackage{bm}
\usepackage{hyperref}
\usepackage{color}

\newcommand{\beq}{\begin{equation}}
\newcommand{\eeq}{\end{equation}}
\newcommand{\bea}{\begin{eqnarray}}
\newcommand{\eea}{\end{eqnarray}}
\newcommand{\e}{{\hat{\mathbf e}}}
\newcommand{\x}{{\hat{\mathbf x}}}
\begin{document}

%
\title{First-passage time of run-and-tumble particles}
\author{L. Angelani\inst{1}\thanks{\email{luca.angelani@phys.uniroma1.it}}, 
R. Di Leonardo\inst{1} \and M. Paoluzzi \inst{1}
%
}                     
%
%
\institute{CNR-IPCF, UOS Roma c/o Dip. di Fisica Universit\`a
  ``Sapienza'', I-00185 Roma, Italy}
%
\date{Received: date / Revised version: date}
%
\abstract{
We solve the problem of first-passage time for run-and-tumble particles in one dimension.
Exact expression is derived for the mean first-passage time in the general case, considering
external force-fields and chemotactic-fields, giving rise to 
space dependent swim-speed and tumble rate.
Agreement between theoretical formulae and numerical simulations is obtained
in the analyzed case studies -- constant and sinusoidal force fields, constant gradient 
chemotactic field. Reported findings can be useful to get insights into very different phenomena involving 
active particles, such as  bacterial motion in external fields, intracellular transport, cell migration, 
animal foraging.
} 
\maketitle
\textcolor{blue}{\href{http://dx.doi.org/10.1140/epje/i2014-14059-4}{DOI: 10.1140/epje/i2014-14059-4}}
\section{Introduction}
\label{intro}
First-passage problems are of great interest in many disciplines,
from  physics and chemistry to engineering and biology \cite{hanggi,redner}.
After the seminal paper of Kramers on kinetic reactions \cite{kramers},
a lot of efforts have been devoted to analyze first-passage processes, or the
related narrow-escape problems \cite{NEP}, 
in a variety of cases, both using theoretical modeling and numerical analysis
\cite{engel,cond,git,schuss,ben,osh,mat}.
The basic question is to find the mean first-passage time (MFPT), i.e. the mean time 
needed for a particles (that can be an atom, molecule, cell, animal or also a signal) 
to reach for the first time a specific site. 
Usually one considers processes in which the particle's motion is described
by Brownian dynamics and the  probability density function (PDF) obeys a
Fokker-Planck equation \cite{risken}. 
This encompasses many real processes involving thermal passive objects, such as 
molecules, atoms or colloidal particles.
In recent years a new kind of systems, characterized by an {\it active} nature,
have been found to describe many interesting situations in real living world, 
from bacterial baths \cite{berg,cates}
to cells migration \cite{cell_mig}
or animals' movements \cite{vz}.
Such  systems are characterized 
by self-propulsion mechanisms inducing a persistent motion, that can be described
by simple random walk models \cite{cod,roman}. 
Among them the run-and-tumble model \cite{cates}
has been found to be a very powerful modeling of real situations, allowing for 
analytic treatments \cite{tail,mart,artw}
and easy implementation in numerical simulations \cite{reic,reich2,ang1,pao}. 
In the simple version of the model, particles perform
a straight line motion at constant speed alternating tumble events 
(assumed to be random Poissonian processes), in which 
particle's orientation randomly changes.
The run-and-tumble model captures the behavior of motile bacteria, 
such as {\it E.coli} \cite{berg,cates},
but is also related to many different  physical processes, 
such as the motion of electrons in metals (Lorentz kinetic model)  
\cite{lorentz,lorentz2}
the propagation of signals in transmission lines (telegraph equation) \cite{gold},
the dynamics instability of microtubules \cite{microtub,microtub2},
the Dirac equation for  relativistic particles in one spatial dimension \cite{dirac_eq,dirac_eq2}.\\
In this paper we analyze the first-passage problem for run-and-tumble particles in 
one dimension. We consider the  general case in which external potentials and 
chemotactic fields are present, inducing a space dependence of particles' speed and
tumble rate. 
A general expression of the mean first-passage time is given
in the general case, and specialized to some interesting case studies -
linear and sinusoidal potentials, constant gradient chemotactic field.
Well known expression valid for Brownian particles is recovered in the limit
of high tumble rate and particle's velocity.

\section{First passage time}
\label{sec:1}

We consider run-and-tumble particles in one dimension.
Denoting with $P_{_R}(x,t)$ and $P_{_L}(x,t)$ the PDFs of 
right-oriented and left-oriented particles,
with $v_{_R}$ and $v_{_L}$ their swim speeds
and with $\alpha_{_R}$ and $\alpha_{_L}$ the tumble rates, 
we can write the continuity equations as follows 
\cite{tail,mart,sch_1993,tail2,LA_epl}
\begin{eqnarray}
\label{eq_r}
\partial_t P_{_R} &=& - \partial_x (v_{_R} P_{_R})
- \frac{\alpha_{_R}}{2} P_{_R} + \frac{\alpha_{_L}}{2} P_{_L} \\  
\label{eq_l}
\partial_t P_{_L} &=& \partial_x (v_{_L} P_{_L})
+ \frac{\alpha_{_R}}{2} P_{_R} - \frac{\alpha_{_L}}{2} P_{_L} 
\end{eqnarray}
We treat here the very general case in which speeds
$v_{_{R,L}}(x)$ and tumble rates $\alpha_{_{R,L}}(x)$ are space dependent quantities,
for example due to the presence of external force fields or chemotactic fields.
We can write the equations for the total PDF
$P(x,t)=P_{_R}(x,t)+P_{_L}(x,t)$ and the current 
$J(x,t)=v_{_R}(x) P_{_R}(x,t) - v_{_L}(x) P_{_L}(x,t)$ as
\beq
\partial_t P = -\partial_x J 
\label{eq_P}
\eeq
and 
\bea
 \partial_t  J  = &-& (v_{_R}+v_{_L})\ \partial_x \left(\frac{v_{_R} v_{_L}}{v_{_R}+v_{_L}} P \right) \nonumber \\
&+& \frac{P}{2} (\alpha_{_L} v_{_R} - \alpha_{_R} v_{_L}) 
-  (v_{_R}-v_{_L})\ \partial_x J  \nonumber \\
&-& \frac{J}{2} \left[  \alpha_{_R} + \alpha_{_L} + (v_{_R}+v_{_L})\ \partial_x 
\left(\frac{v_{_R}-v_{_L}}{v_{_R}+v_{_L}}\right)\right]
\label{eq_J}
\eea
where, for the sake of simplicity, the dependence on space and time  
of the different quantities has not explicitly indicated.
We want to determine the mean first passage time, that is the mean time
needed for a particle starting its motion at $x\!=\!0$ to reach for the first time
the position $L$.
In order to solve the problem we consider
initial conditions
$P_{_R}(x,0)\!=\! \delta(x)$, $P_{_L}(x,0)\!=\! 0$
and  
reflecting/absorbing boundary conditions respectively at $x\!=\!0$ and $x\!=\!L$:
$J(0,t)\!=\! 0$ and $P_{_L}(L,t)\!=\!0$, i.e.
$ J(L,t)\! =\! P(L,t) v_{_R}(L) $.
By noting that the survival probability that the particle has not yet been absorbed at time $t$
is $\mathbb{P}(t)=\int_0^L dx P(x,t)$ and the probability density 
of the first-passage time is $\varphi(t) = -\partial_t \mathbb{P}$,
the MFPT $\tau=\int dt \ t \ \varphi (t)$ can be written as 
\beq
\tau = \int_0^L dx  \ Q(x)
\label{tau}
\eeq
where 
\beq
Q(x) = \int_0^\infty dt\ P(x,t)
\eeq
From Eq.s (\ref{eq_P},\ref{eq_J}) the quantity $Q(x)$ 
obeys the following equation
\bea
&-& (v_{_R}+v_{_L})\ \partial_x \left(\frac{2 v_{_R} v_{_L}}{v_{_R}+v_{_L}} Q \right) \nonumber 
 +  (\alpha_{_L} v_{_R} - \alpha_{_R} v_{_L}) \ Q  \\
&=& \alpha_{_R} + \alpha_{_L} + (v_{_R}+v_{_L})\ \partial_x \left(\frac{v_{_R}-v_{_L}}{v_{_R}+v_{_L}}\right)
- 2 v_{_L} \delta(x)
\label{eq_Q}
\eea
with boundary conditions $Q(L)v_{_R}(L)=1$.
It is worth noting that the above equation is similar to that obtained
for the PDF of a run-and-tumble particle in the stationary case \cite{sch_1993,artw,tail}. 
We have then reduced the problem of MFPT to the search of a solution 
of a stationary equation with particular boundary conditions.
We introduce the following quantities
\begin{equation}
A(x) =  \exp \left\{ \int_0^x dy \ \frac{\alpha_{_L} v_{_R} - \alpha_{_R} v_{_L}}{2 v_{_R}v_{_L}}
\right\}
\label{Ax}
\end{equation}
and 
\begin{equation}
B(x) =  \int_x^{L} \frac{dy}{A(y)} \ 
\left[ \frac{\alpha_{_R} + \alpha_{_L}}{v_{_R}+v_{_L}} +
\partial_y \left( \frac{v_{_R}-v_{_L}}{v_{_R}+v_{_L}}  \right)
\right] 
\label{Bx}
\end{equation}
The solution of Eq.(\ref{eq_Q}) can be written as
\bea
Q(x) &=&  \ \frac{v_{_R}(x)+v_{_L}(x)}{2 v_{_R}(x)v_{_L}(x)} \ A(x) \nonumber\\
&\times& \left[ 
\frac{2v_{_L}(L)}{v_{_R}(L)+v_{_L}(L)} 
\frac{1}{A(L)} 
+ B(x) \right]
\label{Qx}
\eea
The  mean first passage time $\tau$ is finally obtained from Eq. (\ref{tau}).\\
Some interesting limits are:\\
{\it i})\ 
The free particles case 
(absence of  external force fields and chemotactic fields), 
i.e. $v_{_{R,L}}(x)\!=\!v$ and $\alpha_{_{R,L}}(x)\!=\!\alpha$,
leads to 
\begin{equation}
\tau_{free} =\frac{L}{v} +
\frac{L^2}{2D_0}
\label{tau_free}
\end{equation}
where $D_0=v^2/\alpha$.
It is worth noting that the MFPT can be written as the sum of 
a  pure ballistic term plus a  diffusive term.\\
{\it ii})\ 
In the case of no tumbling, $\alpha_{_{R,L}}\!=\! 0$, we get
\begin{equation}
\tau_{no-tumble} = \int_0^{L} \frac{dx}{v_{_R}}
\label{tau_not}
\end{equation}
which is the time a particle with space-dependent velocity $v_{_R}(x)$ takes
to go form $0$ to $L$.\\

In the following we will consider  the two cases in which 
particles are immersed in an external potential field (Section 3)
or in a chemoattractant field (Section 4).
A few case studies will be analyzed, deriving theoretical expressions and comparing them 
with numerical simulations. 
The latter are performed considering run-and-tumble particles
which obey the 1D version of the Eq. of motion 
$\partial_t {\bf r} = v\  \e + \mu \  {\bf f}$ 
, with $\e$ the unit vector indicating the direction of motion of the particle,
$\mu$ the particles mobility and  $f(x)$ the external force.
The direction $\e$ is updated with rate 
$\alpha-\gamma \  \e \cdot \nabla c$ ($c(x)$ is the chemotactic field)
randomly choosing, in a uniform way,  the new one -- $\e=\pm \x$ in 1D
(see the following sections for details).

\section{External forces}
\label{sec:2}

We first analyze the case in which particles feel an external potential field and
chemotactic effects are absent,
i.e. the tumbling  rates of right-oriented and left-oriented particles
can be considered equal and uniform
\begin{equation}
\alpha_{_{R,L}}(x)=\alpha
\end{equation}
and the speeds are given by
\begin{equation}
\begin{split}
v_{_R} (x) &= v + \mu \ f (x)\\  
v_{_L} (x)&= v - \mu \ f (x)
\label{v}
\end{split}
\end{equation}
with $v$ the free-speed  of the  particles, $\mu$ the particles mobility and 
$f(x)=-\partial_x V(x)$ the force due to an external potential $V(x)$.
In such a case, Eq.s (\ref{Ax},\ref{Bx},\ref{Qx}) become:
\begin{eqnarray}
A(x) &=&  \exp \left\{ \int_0^x dy\  \frac{\mu f(y)}{D(y)} \right\} \\
B(x) &=&  \int_x^{L} dy \ (\alpha+  \mu\ \partial_y f)\  [v A(y)]^{-1} \\
Q(x) &=&  \ \frac{vA(x)}{\alpha D(x)} \left[ \frac{v-\mu f(L)}{v A(L)} + B(x) \right] 
\end{eqnarray}
where $D(x)$ is the local diffusion coefficient $D(x) = [v^2-\mu^2 f^2(x)] / \alpha$.\\
It is worth noting that the Brownian case is obtained in the limit 
$\alpha , v \to \infty$ with constant $D_0=v^2/\alpha$,
leading to the well known expression of the mean escape time for Brownian particles \cite{risken,zwan}
\begin{equation}
\tau_{_B} = \frac{1}{D_0} \int_0^{L} dx \ e^{-\beta V(x)}\ 
\int_x^{L} dy \ e^{\beta V(y)}\ 
\label{tau_br}
\end{equation}
where $\beta = \mu / D_0= \mu \alpha /v^2$.\\
We consider here two kinds of external potentials.
The first one is a linear potential
\begin{equation}
V(x)=f_0 \ x
\label{v1}
\end{equation}
where $f_0= \Delta /L$ ($\Delta$ is the barrier height).
In this case the mean escape time $\tau$ has an explicit expression
\begin{equation}
\tau = \frac{v (v+\mu f_0)}{\alpha (\mu f_0)^2} \left( e^{\mu f_0 L/D} - 1 \right) - \frac{L}{\mu f_0}
\label{tau_lin}
\end{equation}
where $D=(v^2-\mu^2 f_0^2)/\alpha$.
In the free limit, $f_0 \to 0$, one has $\tau \!=\! \tau_{free}$ - see Eq.(\ref{tau_free}) - 
while for $ f_0 \to v/\mu$ one has $D \to 0$ and the MFPT diverges as 
$\tau \simeq (2/\alpha) \exp (vL/D )$.
In Fig. 1 (upper panel) the MFPT $\tau$ is reported as a function of $f_0$ for different values
of $L$ \cite{nota1}.
Theoretical predictions 
are in agreement with numerical simulations. 

\begin{figure}
\resizebox{0.5\textwidth}{!}{%
  \includegraphics{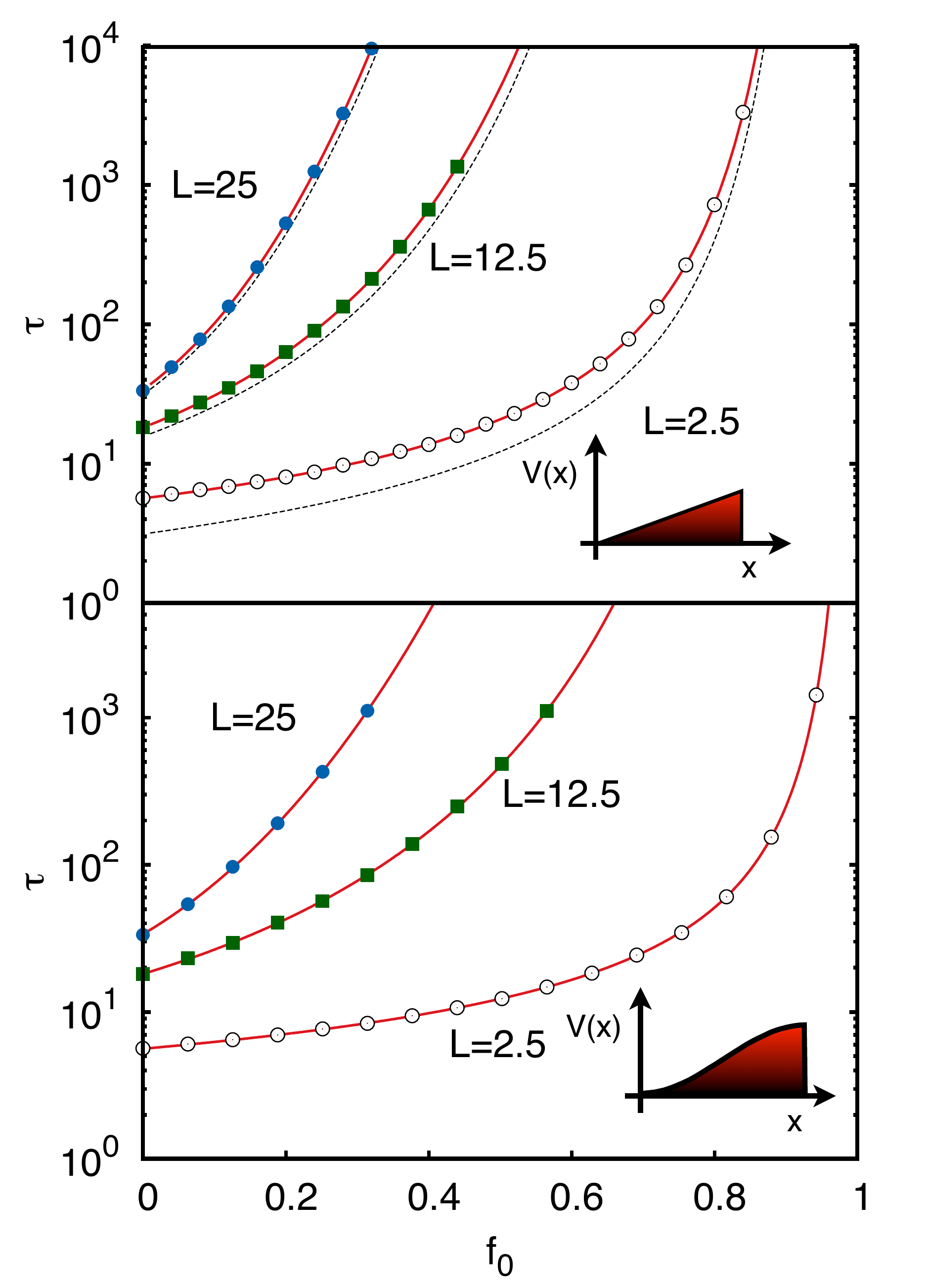}
}
\caption{Mean first-passage time $\tau$ of run-and-tumble particles in external force fields
as a function of field strength.
Upper panel refers to linear potential, 
Eq. (\ref{v1}), while lower panel
to sinusoidal potential,  Eq. (\ref{v2}). 
Lines are theoretical predictions and symbols are from numerical simulations. 
Three different values of box length $L$ are reported.
In the upper panel dashed-lines correspond to the Brownian limit, Eq. (\ref{brl}).
Quantities are expressed in reduced units \cite{nota1}.}
\label{fig:1}       
\end{figure}
%

\noindent In the Brownian limit, $\alpha , v \to \infty$ at constant $D_0 = v^2/\alpha$, one has
\begin{equation}
\tau_{_B} = \frac{D_0}{(\mu f_0)^2} \left( e^{\mu f_0 L/D_0} - 1 \right) - \frac{L}{\mu f_0}
\label{brl}
\end{equation}
We observe that $\tau_{_B} < \tau$, indicating that in the Brownian limit particles
take less time to reach absorbing boundaries. It is worth noting, however, 
that the comparison is made considering equal diffusivities,
and this does not imply that in real situation, at room temperature, 
thermally activated particles are faster than self-propelling ones.
Indeed, just to make an example, considering parameters' values suitable for {\it E.coli}  
($v\!\simeq\!30\ \mu$m s$^{-1}$, $\mu\!\simeq\!60\ \mu$m pN$^{-1}$ s$^{-1}$, $\alpha\!\simeq\!1$ s$^{-1}$) one has
that thermal diffusivity at room temperature is about four order of magnitude less 
than the self-propelling one $D\simeq 10^3 \ \mu$m$^2$ s$^{-1}$ , or, in other words, in order to have the same
diffusivity one has to consider a temperature of $10^6$ K.

\vspace{0.3cm}
\noindent As second example we consider a sinusoidal potential 
\begin{equation}
V(x) = \frac{\Delta}{2} \ \left( 1 - \cos{\pi x / L} \right)
\label{v2}
\end{equation}
where $\Delta$ again is the barrier height.
The corresponding force on particles is
$f(x) = - f_0 \ \sin{\pi x / L}$
, where $f_0=\pi \Delta / 2L$.
We can write an explicit expression for the quantity $A$ defined by Eq.(\ref{Ax}) 
\begin{equation}
A(x) = \exp{\left\{
\frac{\alpha L\ 
[W(x)-W(0)]
}{ \pi v \sqrt{1-\eta^2}} \right\}}
\end{equation}
where  $\eta = \mu f_0 /v$ and the quantity $W$ is given by
$W(x) = \tan^{-1} [ \eta \cos(\pi x/L) / \sqrt{1-\eta^2} ]$.
The MFPT $\tau$ can be obtained by numerical integration of 
Eq.(\ref{tau}), by using Eq.s (\ref{Bx},\ref{Qx}). 
In Fig. 1 (lower panel) the MFPT $\tau$ is reported as a function of $f_0$ for different values
of $L$ \cite{nota1}.
It is worth noting that, as already stated above, 
one observes a divergence of the escape time $\tau$ when the external force equals the self propelling one $v/\mu$ ($f_0 \to 1$ in reduced unit).
Indeed, due to the finite velocity of run-and-tumble particles, they can be easily confined by 
strong enough external fields, inducing interesting trapping phenomena, 
recently investigated in random environments, such as speckle fields \cite{pao2}
or random distribution of obstacles \cite{reich2,chep}.
\vspace{0.5cm}


\section{Chemotaxis}
\label{sec:3}

We now analyze the effects of chemotactic fields.
Bacteria, such as {\it E.coli}, are able to sense the environment and,
through a chemical internal circuit,  modify their tumble rate 
in order to achieve a net preferential motion towards high nutrient concentration
\cite{berg}. The adopted strategy consists in evaluating changes in ambient 
chemical concentration performing a time-integral of concentration $c$ 
filtered by a suitable kernel function \cite{berg}.
A simplified expression of the tumble rate $\alpha$ as a function of chemotactic field 
$c(x)$ is obtained in the limit of weak concentration gradient,
giving rise to space dependent right and left tumble rates
\begin{equation}
\begin{split}
\alpha_{_R} (x)&= \alpha - \gamma v \ \partial_x c (x)\\ 
\alpha_{_L} (x)&= \alpha + \gamma v \ \partial_x c (x)
\label{alpha}
\end{split}
\end{equation}
where $\alpha$ is the tumble rate in the absence of chemoattractants and  
$\gamma$ the strength of the particles reaction to the chemicals
\cite{cates,sch_1993}.
It is worth noting, however, that modeling chemotaxis through spatially varying
tumble rates could be a quite crude approximation of the mechanism
adopted by real organisms, which instead involve more complex time integration \cite{cates}.
By considering force-free particles
\begin{equation}
v_{_{R,L}}(x)= v
\end{equation}
we have the following expression for the quantities $A,B$, and $Q$
\begin{eqnarray}
A(x) &=&  \exp \left\{ \int_0^x dy\  \gamma \ \partial_y c(y) \right\} \\
B(x) &=&  \int_x^{L} dy \ \alpha \  [v A(y)]^{-1} \\
Q(x) &=&  \ \frac{A(x)}{v} \left[ \frac{1}{A(L)} + B(x) \right] 
\end{eqnarray}
and the MFPT takes the form
\beq
\tau = \int_0^L \frac{dx}{v}\ 
e^{\gamma c(x)} 
\left[ 
e^{-\gamma c(L)} 
+ \frac{\alpha}{v} \int_x^L dy\ 
e^{-\gamma c(y)}
\right]
\eeq
\\
We consider here the simple case of constant gradient concentration:
\begin{equation}
c(x)=c_0 \ x
\end{equation}
The MFPT reads
\beq
\tau = \frac{L}{\gamma c_0 D_0} - 
\frac{1}{\gamma^2 c_0^2 D_0}(1-\frac{v \gamma c_0}{\alpha})
(1-e^{-\gamma c_0 L})
\label{tau_lin}
\eeq
\begin{figure}
\resizebox{0.5\textwidth}{!}{%
  \includegraphics{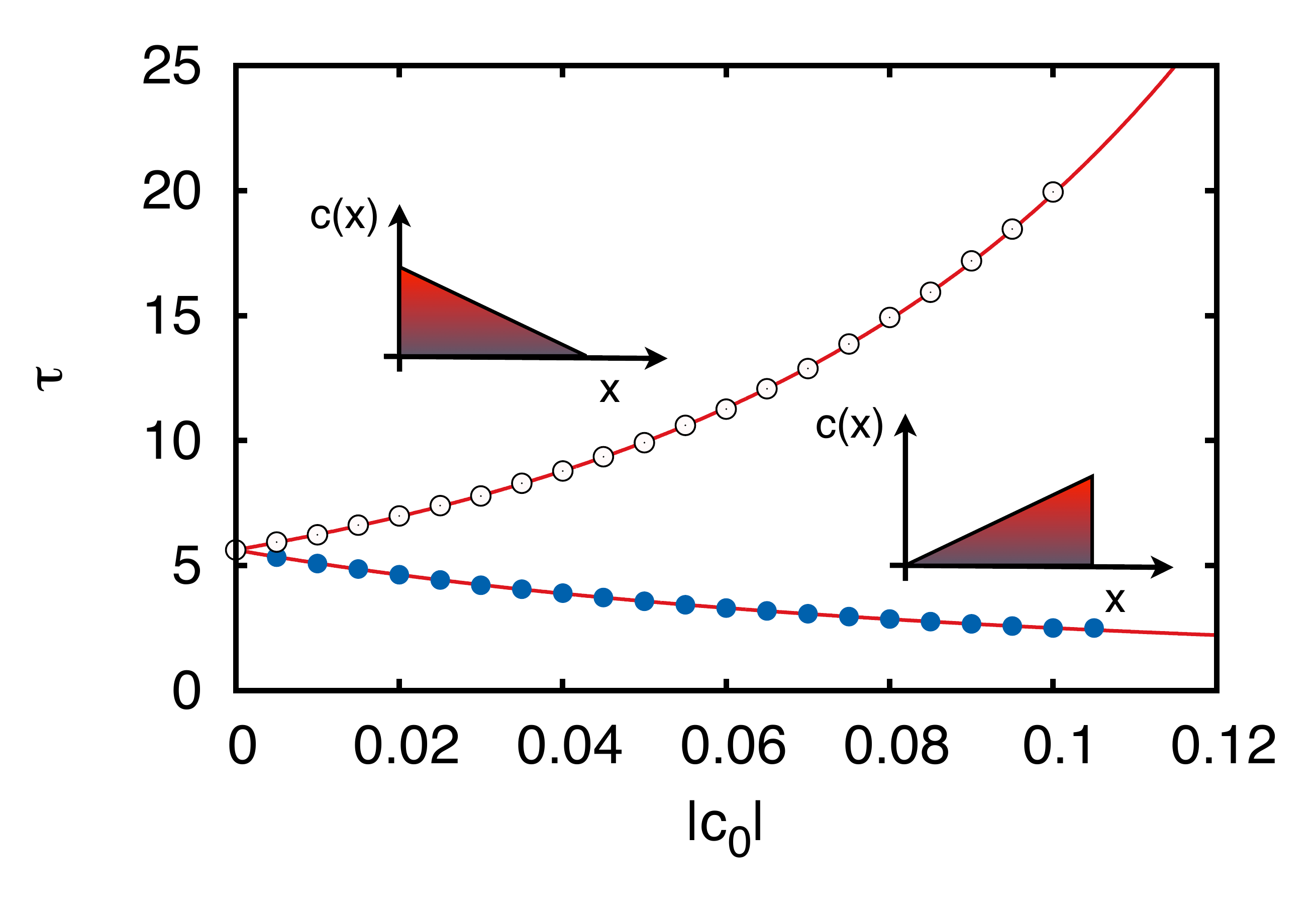}
}
\caption{Mean first-passage time $\tau$ of run-and-tumble particles in chemotactic fields
as a function of field strength.
The two curves refer to the cases of chemoattractant boundary, $c_0>0$ (full symbols)
and chemorepellent boundary, $c_0<0$ (open symbols). 
Lines are theoretical predictions and symbols are from numerical simulations. 
Quantities are expressed in reduced units \cite{nota1}.}
\label{fig:2}       
\end{figure}
%
%
For $c_0>0$ the maximum chemoattractant concentration is at the absorbing 
boundary $L$, resulting in a net preferential swimming towards such a point.
For $c_0<0$ the maximum is at $x\!=\!0$ and reaching the boundary is made 
more difficult by the chemotactic field.
For weak concentration gradient one has:
\beq
\tau \simeq \tau_{free} - c_0 \frac{\gamma L^2}{2v} \left( 1+ \frac{L\alpha}{3v}\right)
\eeq
which, in the absence of chemicals, $c_0=0$, reduce to the free case $\tau \! =\!\tau_{free}$.
In Fig. 2 the MFPT $\tau$, Eq. (\ref{tau_lin}), is reported  as a function of $|c_0|$ for the two cases
of positive and negative $c_0$, corresponding, respectively, to chemoattractant
and chemorepellent boundary.\\
We conclude this section discussing 
a different form of chemotaxis,
in which the cell sets its tumble rate considering the absolute attractant concentration 
instead of its temporal changes. This is the case, for example, of mutant bacteria
lacking particular enzymes required to perform temporal comparisons of
concentration \cite{sch_1993}.
If we assume instantaneous sensing
we can write tumble rates as
\begin{equation}
\alpha_{_{R,L}} (x)= \alpha - \tilde{\gamma} \  c (x)
\end{equation}
leading to the following expression for MFPT
\beq
\tau = \tau_{free} - \frac{\tilde{\gamma}}{v^2} \ \int_0^L dx \ x\ c(x)
\eeq
which, in the case of linear concentration  $c(x)=c_0 \ x$, reduces to
\beq
\tau = \tau_{free} - \frac{\tilde{\gamma} c_0 L^3}{3v^2} 
\eeq
However, real cells measure chemical concentrations in a finite time,
and then a more appropriate form of tumble rates in the case of mutant bacteria
is \cite{sch_1993}
\begin{equation}
\begin{split}
\alpha_{_R} (x)&= \alpha - \tilde{\gamma} c(x) + \gamma v \ \partial_x c (x)\\ 
\alpha_{_L} (x)&= \alpha - \tilde{\gamma} c(x) - \gamma v \ \partial_x c (x)
\label{alpha}
\end{split}
\end{equation}
The ratio $\gamma/ \tilde{\gamma}$ is proportional to the average time 
the cell spends to measure chemical concentration.
The MFTP now reads 
\bea
\tau = &\ & \int_0^L \frac{dx}{v}\  
e^{- \gamma c(x)} \nonumber \\
&\times& \left[ e^{\gamma c(L)} 
+  \int_x^L \frac{dy}{v}\ e^{\gamma c(y)} (\alpha-\tilde{\gamma} c(y))
\right]
\eea
and, for constant gradient field $c(x)=c_0 \ x$
\bea
\tau &=& \frac{L}{\gamma c_0 v^2} 
\left[ \frac{\tilde{\gamma}}{\gamma} (\frac{\gamma c_0 L}{2}-1) - \alpha
\right]\nonumber \\ 
&+&
\frac{e^{\gamma c_0 L}-1}{\gamma^2 c_0^2 v^2}
\left[\alpha+v\gamma c_0 +\frac{\tilde{\gamma}}{\gamma} (1 - \gamma c_0 L )\right]
\eea
For weak concentration gradient one has
\beq
\tau \simeq \tau_{free} + c_0 \frac{\gamma L^2}{2v} \left[ 1+ \frac{L}{3v}
( \alpha - 2 \frac{\tilde{\gamma}}{\gamma} )  \right]
\eeq
indicating a reverse taxis ($\tau > \tau_{free}$) when
$\tilde{\gamma}/\gamma < 3v/2L +\alpha/2$, i.e. when the 
average time required for the cell to measure chemical concentration
is long enough \cite{sch_1993}.


\section{Conclusions}
\label{sec:4}

First-passage time problems for active particles ruled by run-and-tumble dynamics are
solved in one dimension, considering generic space-dependent external forces and 
chemotactic fields.
Exact expressions are given for the mean first-passage time and tested
with simulations in simple case studies. 
A diffusion limit is obtained for  high particle's velocity 
and tumble rate at constant diffusivity, where well known expressions 
for Brownian particles are recovered.
Our results can be of interest to many disciplines, allowing a 
better understanding of phenomena involving persistent motion, 
such as  barrier escaping of active  particles \cite{bur,kouma}, 
intracellular transport \cite{bress},
foraging and predation phenomena in the animal kingdom \cite{foraging,redner2,redner3,col_pred}.

\vspace{1.5cm}
{
We acknowledge support from MIUR-FIRB project RBFR08WDBE. 
RDL acknowledges support from ERC starting Grant (No. 307940).
}

%

\begin{thebibliography}{}
\bibitem{nota1}
{Quantities are expressed in internal units, $\alpha^{-1}$ for time, $v/\alpha$ for length, 
$\mu/v$ for forces.}
\end{thebibliography}
%

\end{document}